\documentstyle[aps,epsf]{revtex}
\input epsf.sty

\preprint{UCTP-113-99}

\begin{document}
\draft

\title{The effective potential of composite diquark fields and the
spectrum of resonances in dense QCD}

\author{V.A.~Miransky$^{a}$, 
I.A.~Shovkovy$^{b}$\thanks{On leave of absence from Bogolyubov
         Institute for Theoretical Physics, 252143, Kiev,
         Ukraine.}
and
L.C.R.~Wijewardhana$^{b}$}

\address{
$^{a}$Bogolyubov Institute for Theoretical Physics,
      252143, Kiev, Ukraine\\
$^{b}$Physics Department, University of Cincinnati, 
      Cincinnati, Ohio 45221-0011}

\date{September 27, 1999}
\maketitle

\begin{abstract}
The effective potential of composite diquark fields responsible for
color symmetry breaking in cold very dense  QCD, in which long--range
interactions dominate, is derived.  The spectrum of excitations and the
universality class of  this dynamics are described.    
\end{abstract}

% 11.15.Ex Spontaneous breaking of gauge symmetries
% 12.38.Aw General properties of QCD
% 12.38.-t Quantum chromodynamics
% 26.60.+c Nuclear matter aspects of neutron stars

%%%%%%%%%%%%%%%%%%%%%%%%%%%%%%%%%%%%%%%%%%%%%%%%%%%%%%%%%%%%%%%%%%%%%%

Recently, there has been considerable interest in the study  of the
color superconducting phase of cold dense QCD 
\cite{ARW1,RSSV1,EHS,PR,Son,Hong,HMSW,SchW,PR-new,other} (for recent
reviews, see Ref.~\cite{WR}). The color superconducting quark matter 
may exist in the interior of neutron stars, with baryon number
densities exceeding a few times the normal nuclear density $n_0 \simeq 
0.17~\mbox{fm}^{-3}$. Also, such matter could be created in
accelerators by heavy ion collisions.

The Ginzburg--Landau (GL) effective action method has been extremely
successful in studying ordinary superconductivity of metals \cite{AGD}.
Recently, a similar approach has been utilized in the study of color
superconductivity \cite{EHS,PR}. However, there the effective action
was postulated based on symmetry and renormalization group arguments,
and not derived from the microscopic theory, QCD.

Following the original approach of Gorkov \cite{AGD}, it would be of a
great interest to derive the effective action in color superconductivity
directly from QCD. In this letter, we make a step in realizing this
program and derive the effective  potential for the order parameter of
color superconductivity in cold dense QCD at such high baryon densities
when the fermion pairing in the diquark channel dominates over that in
the chiral one \cite{ARW1,RSSV1} and when long--range interactions
dominate \cite{PR,Son}. For this purpose, we will utilize the method of
Ref.~\cite{Mir}, which was originally used for the derivation of the
effective action in quenched strong--coupling QED$_4$ (see also
Ref.~\cite{MY}) and then was successfully applied to QED$_3$ \cite{Sh},
quenched QED$_4$ in a magnetic field \cite{LNMS}, and to some other
models \cite{Gorb}.

The crucial feature in the dynamics of cold dense QCD, pointed recently
in Refs.~\cite{PR,Son} (see also Ref. \cite{Hong}),
is the presence of the long--range interactions
mediated by the unscreened gluon modes of the magnetic type. This point
essentially distinguishes the dynamics of color superconductivity from
that in the BCS theory of superconductivity in metals. In particular,
this makes the derivation of the effective action in color
superconductivity  more complicated than the derivation of the GL
effective action from the BCS theory.

Our derivation of the effective potential in dense QCD will be based on
the recent analysis of color superconductivity in the framework of the
Schwinger--Dyson (SD) equations \cite{HMSW,SchW,PR-new}. In this way, we
will describe the universality class of the dynamics in cold dense QCD
and, in particular, get insight into the character of the spectrum of
excitations.

As we shall see below, the universality class of the system at hand is
that connected with long--range non--isotropic forces, producing a
bifermion condensate. We will see  that this class resembles (although
does not quite coincide with) that of quenched QED$_4$ in a constant
magnetic field \cite{GMS,Ng}. The scaling law for the order parameter
$X$ in these two models has the following form: 
\begin{equation}
X=\Lambda_{eff}f(z),
\qquad
f(z)\sim \exp\left(-\frac{C}{\sqrt{z}}\right),
\label{Scal}
\end{equation}
and $C$ is some constant. Here $z$ is a generic notation for parameters
of a theory, such as a coupling constant, temperature, the number of
fermion flavors, {\em etc.}. In QED$_4$ in a magnetic field $B$, the
effective  cutoff $\Lambda_{eff}$ is proportional to  $|eB|^{1/2}$ and
$z$ is the QED running coupling at the scale  $|eB|^{1/2}$. In cold
dense QCD, which is of main interest here, $\Lambda_{eff}$ is
proportional to the chemical potential $\mu$ and $z$ is the running
QCD coupling constant $\alpha_s$ at the scale $\mu$.

The critical value $z_c$ is zero both in cold dense QCD and in QED$_4$
in a  magnetic field. This is because in these two models, strong
interactions are provided by the effective dimensional reduction $3+1
\to 1+1$ in the dynamics of fermion pairing \cite{AGD,GMS}.

One should expect that the long--range interaction in dense QCD leads to
the existence of an infinite number of resonances in different channels.
In particular, as we will see, there is indeed an infinite number of
resonances in the channel with the quantum numbers of the
Nambu--Goldstone (NG) bosons. It will be shown that this in turn leads
to a rather unconventional form of the effective potential: it is a
multibranched function of the bifermion condensate and has an infinite
number of local minima. It reduces to the conventional Coleman--Weinberg
potential \cite{CW} only in the vicinity of the global minimum.  

So, let us consider dense QCD with two light flavors in the chiral
limit. The Lagrangian density reads  
\begin{equation}
{\cal L}_{QCD}=\bar{\psi}\left( i\gamma^{\mu} D_{\mu} +\mu
\gamma^{0}\right)\psi -\frac{1}{2}Tr\left( F_{\mu\nu}
F^{\mu\nu} \right) +{\cal L}_{gf} +{\cal L}_{FP},
\label{L_QCD}
\end{equation}
where ${\cal L}_{gf}$ and ${\cal L}_{FP}$ are the gauge fixing and the
Faddeev--Popov ghost terms. The covariant derivative is defined in a
usual way, $D_{\mu} =\partial_{\mu}-ig_{s} A^{A}_{\mu} T^{A}$, and
$\mu$ is the chemical potential. 

Below, it will be convenient to work with the eight component
Majorana spinors, 
$\Psi=\frac{1}{\sqrt{2}} \left( \begin{array}{c} \psi \\ 
\psi^{C} \end{array} \right)$ 
where
$\psi^{C}=C\bar{\psi}^{T}$ and $C$ is a charge conjugation
matrix, defined by $C^{-1} \gamma_{\mu}C =-\gamma_{\mu}^{T}$
and $C=-C^{T}$. The fermion part of the Lagrangian density
(\ref{L_QCD}) could be rewritten as follows:  
\begin{equation}
{\cal L}_{fer}=\bar{\Psi}\left(\begin{array}{cc} 
i\gamma^{\mu} D_{\mu} +\mu \gamma^{0} & 0\\
0 & i\gamma^{\mu} \tilde{D}_{\mu} -\mu \gamma^{0}
\end{array}\right)\Psi , \qquad 
\tilde{D}_{\mu} =\partial_{\mu}+ig_{s}A^{A}_{\mu}
\left(T^{A}\right)^{T}.
\label{L_fer}
\end{equation}
In order to derive the one--particle irreducible (1PI) effective action
for the local composite field $\hat\phi_{c}(x) =1/2\varepsilon^{ij}
\varepsilon_{abc}  \bar{\psi}^{i}_{a}(x) \gamma^{5} C \left(
\bar{\psi}^{j}_{b}(x) \right)^{T}$ (here $a$, $b$, $c$ and $i$, $j$ are
the color and flavor indices, respectively), whose vacuum expectation
value defines the order parameter in the theory, we need to consider the
corresponding generating functional,  
\begin{equation} 
iW(J_c) =\ln\int d\Psi d\bar{\Psi} dA_{\mu}
\exp\left[i\int d^4 x\Bigg({\cal L}_{QCD} + 
\frac{1}{2}J_c \varepsilon^{ij} \varepsilon_{abc}
\bar{\psi}^{i}_{a} \gamma^{5} C  \left(\bar{\psi}^{j}_{b}
\right)^{T}+c.c.\Bigg)\right]. 
\label{gen-fun}  
\end{equation} 
When the functional $W(J_c)$ is known, the calculation of the
effective action (potential) of interest reduces to performing
the Legendre transform with respect to the external source
$J_c$,  
\begin{equation}
\Gamma(\phi_{c})=W(J_c)-\int d^4 x \left[J_c(x)\phi_{c}(x)
+c.c.\right], \label{eff-act} 
\end{equation} 
where $\phi_{c}(x)= \langle 0|\hat\phi_{c}(x)|0\rangle_{J}$,
and the subscript $J$ implies that $\phi_{c}(x)$ is related to
the theory with a source $J_c$. It is assumed that the
source $J_c$ in Eq.~(\ref{eff-act}) is the function
of the field $\phi_{c}$, obtained by inverting the
expression,
\begin{equation}
\frac{\delta W}{\delta J_c(x)}=\phi_{c}(x). \label{del-W}
\end{equation}
For the purposes of calculating the effective potential of the  field
$\phi_{c}$, it is sufficient to restrict ourselves to the  case of a
constant (in space--time) external source, $J_c(x) =\mbox{Const}$. In
addition, using the freedom of global color transformations, it is
always possible to fix the orientation  of the source in the color space
along the third direction, {\em i.e.}, $J_1=J_2=0$ and $J_3\equiv j\neq
0$. Finally, the baryon symmetry allows us to choose $j$ to be real.

In the theory with the external source, the inverse of the bare
fermion  propagator reads 
\begin{equation}
G_{0}^{-1}=-i\left(\begin{array}{cc} 
\hat{p} +\mu \gamma^{0} & J\\
\gamma^{0}J^{\dagger}\gamma^{0} & \hat{p} -\mu \gamma^{0}
\end{array}\right), \qquad
J^{ij}_{ab} =j\varepsilon^{ij} \varepsilon_{ab3} \gamma^{5}.
\label{G_0}
\end{equation}
Upon neglecting the wave function renormalizations 
\cite{Son,Hong,HMSW,SchW,PR-new}, the inverse of the full
fermion propagator, $G^{-1}$, would be the same as that in
Eq.~(\ref{G_0}) but with $J^{ij}_{ab}$ replaced by
$\Sigma^{ij}_{ab}(p)=\Delta(p) \varepsilon^{ij} 
\varepsilon_{ab3} \gamma^{5}$. By inverting it, we obtain the
following expression for the fermion propagator: 
\begin{eqnarray} 
G &=&i \left(\begin{array}{cc}
R_{1}(p)^{-1} & -\left(\hat{p}+\mu\gamma^{0}\right)^{-1}
\Sigma R_{2}(p)^{-1}\\
-\left(\hat{p}-\mu\gamma^{0}\right)^{-1}
\gamma^{0}\Sigma^{\dagger}\gamma^{0} R_{1}(p)^{-1} &
R_{2}(p)^{-1} \end{array}\right),
\label{G}
\end{eqnarray}
where
\begin{eqnarray}
R_{1}(p)&=&
\left(\hat{p}+\mu\gamma^{0}\right)
-\Sigma \left(\hat{p}-\mu\gamma^{0}\right)^{-1}
\gamma^{0}\Sigma^{\dagger}\gamma^{0},\\
R_{2}(p)&=&\left(\hat{p}-\mu\gamma^{0}\right)
-\gamma^{0}\Sigma^{\dagger}\gamma^{0}
\left(\hat{p}+\mu\gamma^{0}\right)^{-1}\Sigma .
\end{eqnarray}

As is clear from the definition of the fermion propagator, $\Delta(p)$
is directly related to the value of the gap in the fermion spectrum in
the color superconducting phase. At the same time, it is also related to
the vacuum expectation value of the diquark field. Indeed, by making use
of its definition, we obtain
\begin{equation}
\phi\equiv \phi_3 
= \varepsilon^{ij} \varepsilon_{ab3} \mbox{~tr}\left[\left( 
G_{12} \right)^{ij}_{ab} \gamma^{5} \right] 
\simeq  -8 i \int\frac{d^4 p}{(2\pi)^4}\frac{
\Delta(p)}{p_0^2-(|\vec{p}|-\mu)^2-\Delta^2}.
\label{phi-del}
\end{equation}
(Note that this expression, up to the change of notations,
$\Delta \to \phi^{+}_{l-}=-\phi^{+}_{r+}$, would remain the same
if the gap ansatz of Refs.~\cite{PR-Yuk,PR-new} is used. In
notation of Ref.~\cite{SchW}, $\Delta \to \Delta_1$.)  Therefore,
if the solution for the full fermion propagator in the problem
with an external source is known and the function $\Delta(p)$ is
presented, from Eq.~(\ref{phi-del}) we could also obtain the
dependence of the diquark field $\phi$ on the source. And, then,
it is straightforward to calculate the generating functional by
integrating the expression in Eq.~(\ref{del-W}),
\begin{equation}    
w(j)\equiv \frac{W(j)}{\int d^4 x}
=\int^{\Delta_{0}(j)} \phi(\Delta_{0})
\frac{dj(\Delta_{0})}{d\Delta_{0}} d\Delta_{0}, \label{w}
\end{equation}
where, by definition, $\Delta_{0}=\Delta(p)|_{p=0}$.

The gap equation was presented in Refs.~\cite{Son,HMSW,SchW,PR-new}. 
There it was also shown that the Meissner effect is of no  importance
for this equation. The further modification of  the gap equation for the
case of a nonzero external source is straightforward,
\begin{equation} 
\Delta(p_4)\simeq j+\frac{2\alpha_{s}}{9\pi} \int_{0}^{p_4} 
\frac{d q_4 \Delta(q_4)} {\sqrt{q_4^2+\Delta_{0}^2}}
\ln\frac{\Lambda}{p_4} +\frac{2\alpha_{s}}{9\pi}
\int_{p_4}^{\Lambda} \frac{d q_4 \Delta(q_4)}
{\sqrt{q_4^2+\Delta_{0}^2}}\ln\frac{\Lambda}{q_4} ,
\label{gap}
\end{equation}
where $\Lambda = (4\pi)^{3/2}\mu/\alpha_{s}^{5/2}$
\cite{SchW,PR-new} and $\alpha_s$ is the QCD running coupling
related to the scale of order $\mu$. This equation, as is easy
to check, is equivalent to the differential equation,
\begin{equation}
p_4 \Delta^{\prime\prime}(p_4)+\Delta^{\prime}(p_4)
+\frac{\nu^2}{4}
\frac{\Delta(p_4)}{\sqrt{p_4^2+\Delta_{0}^2}}=0,
\qquad \nu=\sqrt{\frac{8\alpha_{s}}{9\pi}},
\label{dif-eq}
\end{equation}
along with the infrared (IR) and ultraviolet (UV) boundary conditions,
$\left. p_4 \Delta^{\prime}(p_4) \right|_{p_4=0}=0$ and
$\Delta(\Lambda) =j$, respectively. Notice that the dependence of the
solution on the source appears only through the UV boundary condition.
The value of the source itself could be interpreted as the bare Majorana
mass. As in Ref.~\cite{HMSW}, we solve the differential equation
analytically in two regions $p_4\ll \Delta_{0}$ and $p_4\gg \Delta_{0}$
and, then, match the solutions at $p_4=\Delta_{0}$.

In the region $p_4\ll \Delta_{0}$, the solution
that satisfies the IR boundary condition reads
\begin{equation}
\Delta(p_4)= \Delta_{0}
J_{0}\left(\nu\sqrt{\frac{p_4}{\Delta_{0}}}\right),
\label{sol-IR}
\end{equation}
where $J_{n}(x)$ is the Bessel function. In the other region,
$p_4\gg \Delta_{0}$, the solution, consistent with the UV
boundary condition, is
\begin{equation}
\Delta(p_4)=
B \sin\left(\frac{\nu}{2}\ln\frac{\Lambda}{p_4}\right)
+j \cos\left(\frac{\nu}{2}\ln\frac{\Lambda}{p_4}\right).
\label{sol-UV}
\end{equation}
Now, by matching the solutions and their derivatives at the
point $p_4=\Delta_{0}$, we get two relations,
\begin{eqnarray}
j&=&\Delta_{0} J_{0}(\nu) \cos\left(\frac{\nu}{2}
\ln\frac{\Lambda}{\Delta_{0}}\right)
-\Delta_{0} J_{1}(\nu) \sin\left(\frac{\nu}{2}
\ln\frac{\Lambda}{\Delta_{0}}\right), \label{j-Del} \\
B&=&\Delta_{0} J_{0}(\nu) \sin\left(\frac{\nu}{2}
\ln\frac{\Lambda}{\Delta_{0}}\right)
+\Delta_{0} J_{1}(\nu) \cos\left(\frac{\nu}{2}
\ln\frac{\Lambda}{\Delta_{0}}\right) .\label{B-Del} \
\end{eqnarray}
The first of them relates the value of the gap in the fermion
spectrum and the strength of the external source, while the
other defines the integration constant $B$ in Eq.~(\ref{sol-UV}).
Now, we can proceed with the calculation of the generating
functional. By using the relation (\ref{phi-del}), we calculate
the vacuum expectation value of the diquark field, 
\begin{equation}
\phi \simeq \frac{4\mu^2}{\pi^2}\int_{0}^{\Lambda}
\frac{d p_4\Delta(p_4)}{\sqrt{p_4^2+\Delta_{0}^2}}
=-\frac{16\mu^2}{\nu^2\pi^2}\Lambda 
\Delta^{\prime}(\Lambda)
=\frac{8\mu^2}{\nu\pi^2}B,
\end{equation}
with $B$ given in Eq.~(\ref{B-Del}). Then, from
Eq.~(\ref{w}) we obtain the generating functional,
\begin{equation} 
w=\frac{\mu^2}{\nu\pi^2}\left[4Bj
+\nu\left(B^2+j^2\right)\right],\label{w-j}
\end{equation}
where $B$ should be considered as a function of $j$,
defined by Eqs.~(\ref{j-Del}) and (\ref{B-Del}). After
performing the Legendre transform, Eq.~(\ref{w-j}) leads to the
effective potential $V(\phi)$ in the following parametric
representation:
\begin{mathletters}
\begin{eqnarray}
V(\Delta_{0})
&=&\frac{\mu^2\Delta_{0}^2}{\nu\pi^2}
\left[2\left( J_{0}^2(\nu) - J_{1}^2(\nu)\right)
\sin\left(\nu\ln\frac{\Lambda}{\Delta_{0}}\right)
\right.
\nonumber\\
&&\left.+4J_{0}(\nu)J_{1}(\nu)
\cos\left(\nu\ln\frac{\Lambda}{\Delta_{0}}\right)
-\nu \left( J_{0}^2(\nu) + J_{1}^2(\nu)\right)
\right] ,\\
\phi(\Delta_{0})&=&
\frac{8\mu^2\Delta_{0}}{\nu\pi^2}
\left[J_{0}(\nu)\sin\left(\frac{\nu}{2}
\ln\frac{\Lambda}{\Delta_{0}}\right)
+J_{1}(\nu)\cos\left(\frac{\nu}{2}
\ln\frac{\Lambda}{\Delta_{0}}\right)
\right].\label{21b}
\end{eqnarray}
\label{Eff-pot}
\end{mathletters}

Let us study the properties of this effective potential. In order to
determine the vacuum expectation value of the diquark condensate,
represented by the composite field $\phi$, we need to know the extrema
of the potential in Eq.~(\ref{Eff-pot}). Thus, we come to the equation
$dV/d\phi=j(\Delta_{0})=0$. By solving it, we obtain an infinite set of
solutions for $\Delta_{0}$,
\begin{equation}
\Delta_{0}^{(n)}=\Lambda\exp\left[-\frac{2}{\nu}\arctan\left(
\frac{J_{0}(\nu)}{J_{1}(\nu)}\right)-\frac{2\pi n}{\nu}\right]
\simeq \Lambda\mbox{e} \exp\left[-\frac{3\pi^{3/2}(1+2n)}
{2^{3/2}\sqrt{\alpha_s}}\right], \quad n=0,1,2,\dots
\label{Gap1}
\end{equation}
which correspond to the following vacuum expectation values of
the diquark field:
\begin{equation}
\phi^{(n)}=(-1)^{n}\frac{8\mu^2\Delta_{0}^{(n)}}{\nu\pi^2}
\sqrt{J_{0}^2(\nu) + J_{1}^2(\nu)}.
\end{equation}
Since $d^2V/d\phi^2|_{\phi^{(n)}}=(\nu\pi/4\mu)^2$,
we conclude that all the extrema are, in fact, minima.

\begin{figure}
\epsfbox{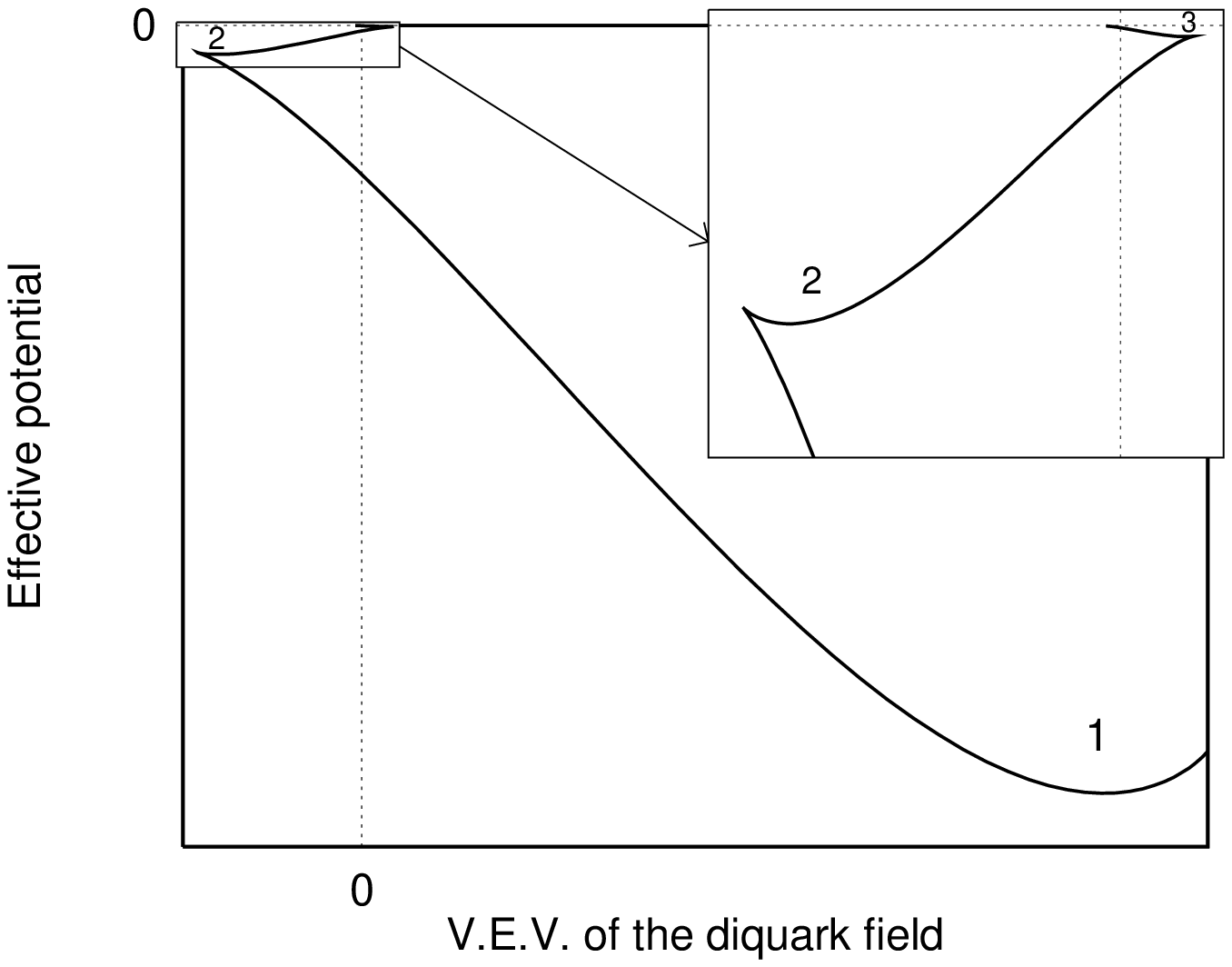}
\caption{An illustration of the multibranched structure of
the effective potential as a function of $\phi$. The enlargement
of the higher (second and third) branches is also shown}
\label{multi-b}
\end{figure}

It is natural to expect that the potential should also have maxima between
those minima. The situation is however more subtle: while, as a function
of the parameter $\Delta_{0}$, the potential does have maxima, it does
not have them as a function of $\phi$. Let us describe this in more
detail. The first derivative of $V$ with respect to $\Delta_{0}$ is zero
at the following maximum points:
\begin{equation}
\Delta_{0max}^{(n)}=\Lambda\exp\left[
-\frac{2}{\nu}\arctan\left(\frac{J_{0}(\nu)}{J_{1}(\nu)}\right)
+\frac{2}{\nu}\arctan\left(\frac{2}{\nu}\right)
-\frac{2\pi n}{\nu}\right]
\simeq \Lambda\exp\left(-\frac{2\pi n}{\nu}\right), 
\quad   n=1,2,\dots 
\end{equation}
However, as is easy to check, the derivative of the potential with
respect to the field $\phi$ at the corresponding $\phi$--points, defined
from Eq.~(\ref{21b}), is nonzero: it is because the derivative of
$\phi$ with respect to $\Delta_{0}$ equals zero there. As one can see
from Fig.~\ref{multi-b}, this property is intimately connected with the
fact that the potential  $V(\phi)$ is a multibranched (multivalued)
function of $\phi$, and these ``maximum" points are sharp turning points
at which different branches of the effective potential merge.

As evident from Fig.~\ref{multi-b}, the physical branch, at which the
potential takes the minimal value for a given value of $\phi$, is the
first branch, at which the global minimum $\phi=\phi^{(0)}$
lies\footnote{Another example of a multibranched potential is connected
with the $\theta$--term in QCD: the QCD effective potential is a
multibranced function of the parameter  $\theta$ \cite{Witten}. The
physical branch is again defined as that with the minimal value of the
potential for a given value of $\theta$.}. It is interesting that the
potential is convex at this branch (we recall that the property of the
convexity of an effective potential follows from general principles of
quantum field theory \cite{Sym}). Moreover, as is seen from
Fig.~\ref{multi-b}, the potential has a fractal structure: after
enlargement, the higher (``small") branches resemble the first
(``large") branch. The whole $n$-th branch shrinks into the limiting
point $\phi=0$ as $n$ goes to infinity.    

As we will show below, this multivaluedness of the potential is
intimately connected with the long--range nature of the interaction in
the model and implies the existence of many different resonances with
the same quantum numbers as the NG bosons.

The global minimum appears at $\phi^{(0)}$. In the vicinity of
this minimum the approximate form of the effective potential is
given by  
\begin{equation} 
V(\phi)\simeq-\left(\frac{\nu\pi}{8\mu}\right)^2 \phi^2
\left[1-\ln\left(\frac{\phi}{\phi^{(0)}}\right)^2\right],
\label{V-app} 
\end{equation} 
{\em i.e.}, in the vicinity of the minimum, it has the form of the
Coleman--Weinberg potential \cite{CW}. The region of validity of this
approximation is given by inequality $\nu\ln(\phi/ \phi^{(0)})\ll 1$,
and, therefore, Eq.~(\ref{V-app}) is a very good approximation for the
potential of the composite field $\phi$ when the coupling is weak or
when the value of the field is close to the minimum. 

Now, let us discuss how the infinite number of minima in the effective
potential (\ref{Eff-pot}) determine the form of the spectrum of the
resonances in the channel with the quantum numbers of NG bosons. It is
well known (see, for example, Ref.~\cite{FGMS}) that, because of the
Ward identities for chiral currents, the SD equation for the dynamical
fermion gap coincides with the Bethe--Salpeter (BS) equation for
corresponding (gapless) NG bosons, which are quark--quark bound states
in the present model. The infinite number of solutions
$\Delta_{0}^{(n)}$ (\ref{Gap1}) for the gap implies that there are
massless states (which would become the NG bosons) in each of the vacua
corresponding to different values of $n$. The genuine, stable, vacuum
is that with $n=0$. What is the fate of the quark--quark bound states
which would be the NG bosons in the false vacua, with  $n=1,2,\dots$?
We will argue below that they become massive,  unstable, particles
there.

In the chiral limit, there are two free parameters in cold dense QCD:
$\Lambda_{QCD}$ and the chemical potential $\mu$, or, equivalently, the
coupling constant $\alpha_{s}(\mu)$ and $\mu$. Let us consider the NG
composites in a false vacuum, with $n$=$n^{(0)}\ge 1$. In that vacuum,
they are massless bound states of fermions with the Majorana mass (gap)
being equal to $\Delta_{0}^{(n_{0})}$. The transition to the genuine
vacuum, with $n=0$, corresponds to increasing the fermion gap,
$\Delta_{0}^{(n_0)}\to \Delta_{0}^{(0)}$, {\em without} changing the
dynamics: the coupling constant $\alpha_{s}(\mu)$ and the chemical
potential $\mu$ remain of course the same. As a result of the increase
of the mass of their constituents, the square of the mass of these bound
states will also increase. Therefore they become massive (apparently,
unstable) composites in the genuine vacuum\footnote{The quenched
strong--coupling QED yields an example of a simpler model with an
effective potential having the form similar to that in
Eq.~(\ref{Eff-pot}) \cite{Mir,MY}. The study of the BS equations in
that model shows that there is indeed an infinite number of resonances
in the channel with the quantum numbers of the NG bosons \cite{FGMS}.
Their masses are nearly equal and are of the order of the fermion
dynamical mass.}. 

Thus we conclude that the global minimum $\Delta_{0}^{(0)}$ of the
effective potential indeed defines the dynamical gap (Majorana mass) of
fermions, and all other minima $\Delta_0^{(n)}$, $n=1,2,\dots$, manifest
the existence of massive radial excitations of NG bosons. Notice that,
because of the Higgs effect, the NG bosons are ``eaten" by the five
gluons, corresponding to the $SU(3)_{c}\to SU(2)_{c}$ breakdown. All the
massive excitations, though, will not be affected by the Higgs mechanism.

In order to determine the spectrum of these massive excitations, one
needs to study the BS equations for massive bifermion bound states in
dense QCD. This problem is beyond the scope of this letter. However, it
is not difficult to estimate their masses: since the fermion gap
$\Delta_{0}^{(0)}$ is essentially the only relevant dimensional
parameter in the pairing dynamics, the masses of these resonances
should be of the order of the fermion gap. The resonances are unstable,
although, they might be rather narrow at high density because the
coupling constant is weak. The presence of such resonances would be a
very clear signature of long--range forces in dense QCD.

Now we come to the description of the universality class of the dynamics
in cold dense QCD. The scaling law for the order parameter is described
by expression (\ref{Scal}) with  $X=\Delta_{0}^{(0)}$, $\Lambda_{eff}
\sim \mu$, and $z=\alpha_s$. The essential singularity at $\alpha_{s}=0$
is provided by long--range forces. Let us discuss the character of these
forces in more detail. 

The gap equation (\ref{gap}) in the absence of the external source can
be rewritten in a different form (see Ref.~\cite{HMSW}), 
\begin{equation}
\Delta(p_4)\simeq \frac{4\alpha_{s}}{9}
\int_{-\infty}^{\infty} \frac{d q}{2\pi}
\int_{-\Lambda}^{\Lambda} \frac{d q_4}{2\pi} 
\frac{\Delta(q_4)}{q^2+q_4^2+\Delta_{0}^2}
\ln\frac{\Lambda}{|q_4-p_4|}  ,
\label{gap-new}
\end{equation}
where the new integration parameter $q$ is the spatial momentum
shifted by the chemical potential, $q=|\vec{q}|-\mu$. Then, it
is easy to show that  Eq.~(\ref{gap-new}) is equivalent to the
following Schr\"{o}dinger equation:
\begin{equation}
\left(-\frac{d^2}{d\tau^2}-\frac{d^2}{dx^2}+\Delta_{0}^2
+U(\tau,x) \right) \Psi(\tau,x)=0,
\label{gap-Schr}
\end{equation}
where 
\begin{eqnarray}
\Psi(\tau,x)&=& \int\frac{d p}{2\pi}
\int\frac{d p_4}{2\pi}
\frac{\Delta(p_4)}{p^2+p_4^2+\Delta_{0}^2}
e^{ip_4\tau-ipx},\label{psi-def}\\
U(\tau,x) &=& -\frac{2\alpha_{s}}{9\pi} 
\delta(x)\int_{-\Lambda}^{\Lambda} d p_4
\ln\frac{\Lambda}{|p_4|} e^{ip_4 \tau} 
=-\frac{2\alpha_{s}}{9\pi |\tau|}
\left[\pi+2\mbox{~si}(|\Lambda\tau|)\right]\delta(x).
\label{U-def}
\end{eqnarray}
Here $\mbox{si}(z)=-\int_{z}^{\infty}dt\sin(t)/t$ is the sine integral
function. So, we see that the problem reduces to the Schr\"{o}dinger
equation (\ref{gap-Schr}) with a  non--isotropic interaction potential
presented in Eq.~(\ref{U-def}). This interaction is short range in the
spatial direction, $x$, and long--range in the (imaginary) time
direction, $\tau$ [notice that $\mbox{si}(z)\simeq -\cos(z)/z$ as $z\to
+\infty$]. It is the latter long--range portion of the interaction that
is responsible for the particular scaling law of the order parameter as
in Eq.~(\ref{Scal}).

In some respects, the dynamics in cold dense QCD is similar to the
dynamics in quenched QED$_4$ in a constant magnetic field \cite{GMS,Ng}.
Indeed, in both these models the dimensional reduction $3+1\to 1+1$ in
the dynamics of fermion pairing takes place. This feature and long--range
interactions lead to the same scaling law (\ref{Scal}) for the order
parameter, which is qualitatively different from that of the BCS type.
The form of the effective potentials in these two models is also similar
(compare Eq.~(\ref{Eff-pot}) with the expression for the potential in
Ref.~\cite{LNMS}). At the same time, the universality class of the
system at hand is somewhat different from that in QED$_4$ in a magnetic
field: in that model, the dynamics is provided by relativistic
Coulomb--like forces. As is clear, the difference appears due to the
explicit breakdown of Lorentz boost transformations in dense QCD by a
nonzero chemical potential [notice that there is the (1+1)--dimensional
Lorentz symmetry in QED$_4$ in a magnetic field]. 

In conclusion, in this letter, we have taken the first step in deriving
the effective action in color superconductivity of the dense quark
matter directly from QCD. In particular, we have derived the 1PI
effective potential for the order parameter responsible for color
symmetry breaking. In this derivation, we used the common assumption
that the baryon density is high enough, so that the fermion pairing in
the color antitriplet channel dominates over that in the chiral one. 

The crucial feature in the dynamics of cold dense QCD is the long--range
interactions mediated by the unscreened magnetic gluon modes
\cite{PR,Son}. Because of these long--range interactions, we argue that
the system belongs to a universality class that is close to (but not
quite the same as) that of quenched QED$_4$ in an external magnetic
field \cite{GMS,Ng}.

We also argue that the spectrum of the diquark resonances with the same
quantum numbers as those of NG bosons consists of a very large (infinite
in our approximation) number of states with masses of order of the
fermion gap. Even though these resonances are unstable, they might be
relatively narrow at sufficiently high density of quark matter. We
believe that the presence of such resonances would be a clear signature
of the unscreened long--range forces in dense QCD. It would be worth
studying in detail the properties of these resonances  under the
conditions produced in heavy ion collisions: if the color
superconducting phase is ever going to be produced in heavy ion
collisions, the detection of these diquark resonances might be a crucial
piece of information for determining the nature of the phase. In further
studies, it would be interesting to clarify the properties of resonances
in other channels as well as to derive the effective potential in the
case of intermediate densities, when the chiral and the diquark
condensates compete \cite{EHS,PR}.

{\bf Acknowledgments}.
V.A.M. thanks V.P.~Gusynin for useful discussions. The work of
I.A.S. and L.C.R.W. was supported by U.S. Department of Energy 
Grant No. DE-FG02-84ER40153.

\end{document}